\documentclass[onecolumn]{emulateapj}
\slugcomment{{\it Accepted to ApJ Nov 26, 2007.}}

\def\ni{\noindent}

\def\v{\mathbf{v}}

\usepackage{epsfig}

\begin{document}

\shortauthors{Chiang}
\shorttitle{Vertical Shearing Instability}

\title{Vertical Shearing Instabilities in Radially Shearing Disks:\\
The Dustiest Layers of the Protoplanetary Nebula}

\author{E.~Chiang\altaffilmark{1}}

\altaffiltext{1}{Center for Integrative Planetary Sciences,
Astronomy Department,
University of California at Berkeley,
Berkeley, CA~94720, USA}


\email{echiang@astro.berkeley.edu}

\begin{abstract}
Gravitational instability of a vertically thin, dusty sheet near the midplane
of a protoplanetary disk has long been proposed as a way of forming
planetesimals. Before Roche densities can be achieved, however, the dust-rich
layer, sandwiched from above and below by more slowly rotating dust-poor
gas, threatens to overturn and mix by the Kelvin-Helmholtz instability (KHI).
Whether such a threat is real has never been demonstrated: the Richardson
criterion for the KHI is derived for 2-D Cartesian shear flow and does not
account for rotational forces. Here we present 3-D numerical simulations
of gas-dust mixtures in a shearing box, accounting for the full suite
of disk-related
forces: the Coriolis and centrifugal forces, and radial tidal gravity.
Dust particles are assumed small enough to be perfectly entrained in gas; the
two fluids share the same velocity field but obey separate continuity
equations.
We find that the Richardson number $Ri$ does not alone determine stability.
The critical value of $Ri$ below which the dust layer overturns and mixes
depends on the height-integrated metallicity $\Sigma_d/\Sigma_g$
(surface density ratio of dust to gas).
Nevertheless, for $\Sigma_d/\Sigma_g$ between one and five times solar,
the critical $Ri$ is nearly constant at $\sim$0.1.
Keplerian radial shear stabilizes those
modes that would otherwise disrupt the layer at large $Ri$. 
If the height-integrated metallicity is at least $\sim$5 times greater than
the solar value of 0.01, then midplane dust densities
can approach Roche densities. Such an environment might be expected to produce
gas giant planets having similarly super-solar metallicities.
\end{abstract}


\section{INTRODUCTION}
\label{intro}

How do dust grains, known to permeate disks surrounding young stars,
assemble into planets? One proposed stage of growth involves
gravitational instability of a dust-rich layer
at the disk midplane (Safronov 1969;
Goldreich \& Ward 1973). If dust grains are free to
settle vertically out of gas,
midplane dust densities eventually exceed Roche densities, and dust
particles can begin to aggregate by self-gravity.

An objection to this means of forming planetesimals is that vertical
velocities of gas may be too high for dust to settle (Weidenschilling 1980).
Even apart from turbulence intrinsic to gas (e.g., turbulence driven
by the magneto-rotational instability), dust layers that are too vertically
thin can suffer from the Kelvin-Helmholtz instability (KHI). The KHI
threatens to manifest because dust-rich gas at the midplane rotates
at a rate different from that of dust-poor gas at higher altitude.
Dust-poor gas experiences greater acceleration from a background
radial pressure gradient, and so its rotation velocity in centrifugal
balance must deviate more strongly from the Keplerian value.\footnote{
This deviation from purely Keplerian flow is analogous to a ``thermal wind''
in the geophysical literature. See Knobloch \& Spruit (1986)
or Chapter 1 of Pedlosky (1979).}
The deviation is smaller for gas laden with dust, since dust adds inertia
but contributes no pressure. Usually it is assumed that the background
radial pressure gradient $\partial P/\partial r < 0$ so that the vertical
shear in the rotation velocity $\partial v_{\phi}/\partial z <0$.

Whether the disk is KH-unstable is commonly assessed using the Richardson
number (Chandrasekhar 1961, page 491):

$$
Ri \equiv \frac{ (g/\rho) \partial \rho/\partial z }{(\partial v_{\phi} /
\partial z)^2 }
$$

\noindent where $g$ is the vertical gravitational acceleration and $\rho$ is
the total mass density. In the limit that perturbations are incompressible,
the numerator is the square of the Brunt-V\"{a}is\"{a}l\"{a} frequency of
buoyant
oscillations, while the denominator is the square of the vertical
shearing rate. The Richardson number measures the amount of work required
to overturn fluid elements that are originally
in hydrostatic equilibrium, against the amount
of free kinetic energy available in the background shear. For purely
Cartesian flow (no rotational forces),

\begin{equation}
Ri < Ri_{\rm crit} = 1/4 \,\,\,\,{\rm is\,\,necessary\,\,for\,\, instability}
\label{onequarter}
\end{equation}

\noindent (Miles 1961; Howard 1961; Chandrasekhar 1961; Drazin \& Reid 1981;
Li et al.~2003).
Despite the fact that (\ref{onequarter}) is formally not a sufficient
criterion, laboratory experiments bear out its usefulness (Tritton 1988).

In the case of our disk, close to the midplane, $g = -\Omega^2 z$ arises
from the tidal gravity of the star, $z$ measures distance from the midplane,
$\Omega$ is the Keplerian angular
frequency, and $\rho = \rho_g + \rho_d$, where $g$ and $d$ denote
gas and dust, respectively. Moreover,
$|\partial \rho_g / \partial z | \ll |\partial \rho_d / \partial z|$
within the thin dust layers of interest to us (see \S\ref{init}).
Thus throughout this paper

$$
Ri = -\frac{\Omega^2 z}{\rho_d + \rho_g} \frac{\partial \rho_d / \partial z }{
\left( \partial v_{\phi} / \partial z \right)^2} \,.
$$

\noindent As dust settles vertically, $\rho_d \sim \Sigma_d / z_d$ increases;
the dust surface density $\Sigma_d$ is assumed constant
while the characteristic scale height $z_d$ of dust decreases.
In the limit that $\rho_d \gg \rho_g$,
$Ri \propto z_d^2$. Criterion (\ref{onequarter}) for instability
therefore places a lower bound on $z_d$ and a corresponding
upper bound on $\rho_d$.
Unfortunately, for conventional solar nebular parameters---in particular
for a height-integrated solar metallicity of
$\Sigma_d / \Sigma_g = 0.01$---the maximum of $\rho_d$ falls short of the
Roche
density by 1--2 orders of magnitude (e.g., Sekiya 1998; Youdin \& Shu 2002).

But is it appropriate to apply criterion (\ref{onequarter})
to a circumstellar disk where the flow is not Cartesian?
One might guess that rotational forces merely modify $Ri_{\rm crit}$
by a factor of order unity, since the Coriolis force and the radial
shear ($d\Omega/dr$) operate on a timescale $\Omega^{-1}$---the same
timescale characterizing vertical Brunt-V\"{a}is\"{a}l\"{a}
oscillations and the vertical shear at $z\sim z_d$.
Yet when Coriolis forces are introduced, as in the
analysis of G\'{o}mez \& Ostriker (2005, hereafter GO), the stability
properties of the flow change dramatically. Numerical simulations by GO,
performed under the assumption that dust is perfectly entrained in gas,
reveal no well-defined threshold for instability. Unstable modes are
detected for $Ri$ as high as 4, with growth rates that diminish with
increasing $Ri$ but that show no sign of vanishing (see their Figures 9 and
10). Similar results obtain in numerical simulations by
Johansen, Henning, \& Klahr (2006,
hereafter JHK), who relax the assumption of perfect coupling between dust
and gas but who, like GO, retain only Coriolis forces and ignore
radial shear. For their ``rocks'' and ``pebbles'' with momentum stopping
times 10--50 $\times$
shorter than $\Omega^{-1}$, the dust distribution evolves to one where
$Ri$ increases from $\sim$1 at the midplane to $\gtrsim 10$ at higher
altitude (JHK, their Figures 4 and 5).

Here we conduct numerical simulations that account
for the full complement of disk-related forces. We include
not only the Coriolis force but also the centrifugal force
and radial tidal gravity. The latter two forces
combine to produce radial shear in rotational equilibrium.
Thus we investigate the stability of doubly shearing
flows: vertically shearing dust layers in near-Keplerian differential
rotation. We expect our results to differ from those
of GO and JHK. The modes emphasized by GO have growth rates
$\lesssim 0.1\Omega$ at $Ri \gtrsim 2$. Keplerian differential rotation,
characterized by a strain rate of $3\Omega/2$, should shear such modes
apart before they have time to amplify.

Indeed this is the conclusion
of Ishitsu \& Sekiya (2003, hereafter IS),
who numerically integrate the linearized
equations of motion, including the full suite of disk-related forces,
to compute the factors by which modes amplify. They find that amplification
factors are sufficiently limited by radial shear that midplane dust-to-gas
density ratios might reach values as high as $\sim$2 (still too low
for gravitational instability, unfortunately). They restrict, however,
their linear analysis to odd-parity modes for which the vertical
velocity $v_z(z) = -v_z (-z)$. G\'{o}mez \& Ostriker (2005) establish
that even-parity modes grow faster and overturn the dust layer more
effectively.
Our simulations do not pre-select for either type of mode. Another difference
between our work and IS is that we concentrate on $Ri= {\rm constant}$
flows, whereas IS employ a simple, analytically tractable dust distribution.
Of course, $Ri={\rm constant}$ is merely a plausible condition to which the
dust-gas mixture might relax. One of the goals of this study is
to assess whether the Richardson number is a good predictor of stability
even when all disk-related forces are accounted for. Finally,
the analysis of IS is linear, while our numerical simulations
enable access to nonlinear phenomena.

Our numerical simulations, like those of GO, assume that dust particles
are perfectly
entrained in gas, i.e., we assume that grain momentum stopping times
$\ll \Omega^{-1}$. This approximation is valid for small particles,
e.g., having sizes $\ll 1$ m at disk radius $r = 1$ AU in a minimum-mass
nebula
(e.g., Weidenschilling 1977). The hydrodynamics code and
initial conditions are described in \S\ref{method}.
Results are presented in \S\ref{res}; there we determine $Ri_{\rm crit}$
for our doubly shearing flows. A summary is provided in \S\ref{sum}.

While our entire study is rooted in the literature
on the Kelvin-Helmholtz instability and its impact on planetesimal formation,
and as such revolves around the Richardson criterion, it was brought to our
attention by the referee that the KHI may not be the only
instability afflicting the dusty midplane.
Baroclinic instabilities, studied in disks by Cabot (1984),
Knobloch \& Spruit (1986), and Arlt \& Urpin (2004) may also be relevant.
These authors study baroclinity specifically in the context
of a non-zero vertical shear. More generally, a baroclinic
flow is one whose isobaric surfaces do not coincide with its isodensity
surfaces. Our flows are baroclinic because of molecular weight gradients:
dust adds to the density but not to the pressure.
Baroclinic instability can strike even when the Richardson
number is large. We will connect our findings to the baroclinic
instability when appropriate; indeed, in \S\ref{cori} we propose that
the instability discovered by GO in their simulations of non-radially-shearing
disks is, in fact, the baroclinic instability discovered by Cabot (1984).
And we will see in \S\ref{varsig} that the Richardson criterion
alone does not determine stability under arbitrary conditions.

\section{METHOD}
\label{method}

\subsection{Dust-Gas Equations in Tight Coupling Limit}
\label{basic_eq}

Gas and pressureless dust are two fluids obeying separate
momentum equations. In an inertial frame,

\begin{eqnarray}
\frac{\partial \v_g}{\partial t} + (\v_g \cdot \nabla) \v_g & = & -\nabla \Phi
+ \frac{\rho_d}{\rho_g} \frac{(\v_d - \v_g)}{t_{\rm stop}} - \frac{\nabla
P}{\rho_g} \label{one}\\
\frac{\partial \v_d}{\partial t} + (\v_d \cdot \nabla) \v_d & = & -\nabla \Phi
- \frac{(\v_d - \v_g)}{t_{\rm stop}} \label{two}
\end{eqnarray}

\ni where $\v$ is velocity,
$t$ is time, and $\Phi$ is the background potential.
Terms containing $t_{\rm stop}$, the momentum stopping time for dust in gas,
account for how the two fluids interact frictionally.

In this paper, we work exclusively in the tight coupling limit where particles
and gas have negligible relative velocity, $|\v_d-\v_g|\ll |\v_g|$.
This corresponds
to the case where dust particles are so small---i.e., their surface area
to mass ratios are so large---that they become entrained in the gas
flow over short times. By computing the difference and appropriately weighted
sum of (\ref{one}) and (\ref{two}) (see Youdin \& Goodman 2005), and
taking the limit $t_{\rm stop} \rightarrow 0$, we find

\begin{eqnarray}
\v_d - \v_g & = & 0 \label{three}\\
\frac{\partial \v}{\partial t} + (\v \cdot \nabla) \v & = & -\nabla \Phi -
\frac{\nabla P}{\rho_p + \rho_g} \label{four}
\end{eqnarray}

\ni where $\v = \v_d = \v_g$.

Furthermore,

\begin{eqnarray}
\Phi & = & -\frac{GM}{\sqrt{r^2 + z^2}} \\
P & = & (\gamma-1) \varepsilon \label{eos} \\
\frac{\partial \varepsilon}{\partial t} + \nabla \cdot (\varepsilon \v) & = &
-P \nabla \cdot \v
\end{eqnarray}

\ni where $G$ is the gravitational constant, $M$ is the central stellar mass
at the origin, $\varepsilon$ is the internal energy density of gas,
and $r$ and $z$ are the cylindrical radius and height.
The gas obeys a polytropic equation of state $P = K \rho_g^{\gamma}$
with constant $K$ and $\gamma$. Finally, dust and gas obey separate
continuity relations

\begin{eqnarray}
\frac{\partial \rho_g}{\partial t} + \nabla \cdot (\rho_g \v) & = & 0
\label{continuity}\\
\frac{\partial \rho_d}{\partial t} + \nabla \cdot (\rho_d \v) & = & 0 \,.
\label{nine}
\end{eqnarray}

\ni Equations (\ref{four})--(\ref{nine}) are the ones we solve in this paper,
in the shearing box approximation (Goldreich \& Lynden-Bell 1965;
Hawley, Gammie, \& Balbus 1995; \S\ref{init}).
They contain the same content as equations (2)--(6) of GO.

In the tightly coupled limit, a
fluid parcel does not change its dust-to-gas ratio in a Lagrangian sense:
$d(\rho_d/\rho_g)/dt = 0$, where $d/dt$ is the convective derivative.
Particles cannot slip away from gas,
and so we cannot expect our simulated flows to relax,
from arbitrary initial conditions, into a unique, steady state of marginal
stability (if such a state actually exists). In reality,
such relaxation occurs over the time it takes dust to settle vertically
relative to gas. Vertical settling times are measured in 
Myrs for micron-sized particles (or $\sim$$10^2$ yr for centimeter-sized
particles) in a minimum-mass nebula. Nevertheless,
our equations do permit dust-rich parcels to settle toward the midplane
if their weight
cannot be supported in vertical hydrostatic equilibrium. 
We are able to evolve dust-gas mixtures over dynamical times,
measured in orbital periods, provided $t_{\rm stop}$ is still shorter.
Our simulations
can thus determine whether a given set of equilibrium conditions
is dynamically
stable. We now specify these equilibrium initial conditions.

\subsection{Initial Conditions: Constant $Ri$ Flows}
\label{init}

Call $r_0$ and $\phi_0 = \Omega_0t$ the radius and azimuth
of a test
particle moving on a circular orbit, where $\Omega_0 = \sqrt{GM/r_0^3}$.
Shifting to axes centered on the test particle and rotating at $\Omega_0$,
we trade the usual cylindrical coordinates $(r,\phi,z)$ for their
shearing sheet counterparts
$(x,y,z)$: $x \equiv r-r_0$, $|x|\ll r_0$, $y = (\phi-\phi_0)r_0$,
$|y| \ll r_0$, and $|z| \ll r_0$. In this rotating frame, the momentum
equation
(\ref{four}) reads

\begin{mathletters}
\label{abc}
\begin{eqnarray}
\frac{\partial v_x}{\partial t} + v_i \frac{\partial v_x}{\partial x_i} & = &
\frac{-1}{\rho_g+\rho_d}\frac{\partial P}{\partial x} + 2\Omega_0 v_y + 2q
\Omega_0^2 x \label{a}\\
\frac{\partial v_y}{\partial t} + v_i \frac{\partial v_y}{\partial x_i} & = &
\frac{-1}{\rho_g+\rho_d}\frac{\partial P}{\partial y} - 2\Omega_0 v_x
\label{b}\\
\frac{\partial v_z}{\partial t} + v_i \frac{\partial v_z}{\partial x_i} & = &
\frac{-1}{\rho_g+\rho_d}\frac{\partial P}{\partial z} - \Omega_0^2 z \label{c}
\end{eqnarray}
\end{mathletters}

\ni where $i = (1,2,3)$,
$(x_1,x_2,x_3) = (x,y,z)$, $(v_1,v_2,v_3) = (v_x,v_y,v_z)$, and
curvature terms of order $v^2/r$ are dropped, following
the shearing sheet approximation.  We introduce the variable
$q = - d\ln \Omega / d \ln r$ in (\ref{a})
because in our simulations we experiment
with $q=0$ (zero radial shear) to connect to GO. For the full problem
with a point-mass potential, $q=3/2$.

In rotational equilibrium, equation (\ref{a}) for radial momentum
balance gives


\begin{equation}
v_y = -q \Omega_0 x + \frac{1}{2\Omega_0(\rho_d+\rho_g)} \frac{\partial
P}{\partial x} \,.
\label{shear}
\end{equation}

\ni Initially,
$\partial P / \partial x = (\partial P / \partial r)_{t=0}$ reflects
how the background nebular pressure changes radially on scales of $r$.
We express $(\partial P / \partial r)_{t=0}$ in terms of a model input
parameter, $v_{\rm max}$:

\begin{equation}
\left( \frac{\partial P}{\partial r} \right)_{t=0} \equiv -2 \rho_g (z,t=0)
v_{\rm max} \Omega_0 \,.
\label{dpdr0}
\end{equation}

\ni The velocity $v_{\rm max}$ represents the difference between the Keplerian
rotation rate and the sub-Keplerian rotation rate of pressure-supported gas;
it
measures the maximum possible difference in rotational
velocity between the dust-rich midplane and dust-free gas at higher
altitude. For typical nebular parameters,
it is approximately $20$ m/s and nearly constant with $r$.

Further defining the local dust-to-gas ratio $\mu \equiv \rho_d / \rho_g$,
we rewrite (\ref{shear}) as

\begin{equation}
v_y (x,z,t=0) = -q \Omega_0 x - \frac{v_{\rm max}}{1+\mu(z)}\,.
\label{vy}
\end{equation}

\ni In each of our simulations, we hold $v_{\rm max}$ fixed for simplicity.
Then the vertical shear, $\partial v_y / \partial z$, is non-zero only
when $\partial \mu / \partial z$ is non-zero. Because $\mu$ varies
rapidly with $z$ for the vertically thin dust layers of interest to us,
approximating $v_{\rm max}$ as constant introduces negligible error.

All our simulations investigate the stability of constant $Ri$ flows.
The condition $Ri = {\rm constant}$ yields $\mu(z)$:

\begin{eqnarray*}
Ri & = & -\frac{\Omega_0^2 z}{\rho_d + \rho_g} \frac{\partial \rho_d / \partial
z }{ \left( \partial v_y / \partial z \right)^2} \\
 & = & -\frac{\Omega_0^2}{v_{\rm max}^2} \frac{ (1+\mu)^3 z}{\partial \mu /
\partial z}
\end{eqnarray*}

\ni which integrates to

\begin{equation}
\mu(z) = \left[ \frac{1}{1/(1+\mu_0)^2 + (z/z_d)^2} \right]^{1/2} - 1
\label{mu}
\end{equation}

\ni where $\mu_0 \equiv \mu(z=0)$ is a model input parameter and

$$
z_d \equiv \frac{Ri^{1/2} v_{\rm max}}{\Omega_0}
$$

\ni is a characteristic dust height. Equation (\ref{mu}) implies that dust
extends to a maximum height

$$
z_{\rm max} = z_d \,\frac{ \mu_0^{1/2} (2+\mu_0)^{1/2}}{1+\mu_0} \,.
$$

Finally, vertical hydrostatic equilibrium gives $\rho_g (z)$:

\begin{equation}
\frac{1}{\rho_g + \rho_d} \frac{\partial P}{\partial z} = - \Omega_0^2 z
\label{hydro}
\end{equation}

\ni which, using (\ref{mu}),
can be solved analytically for $\rho_g$. The expressions, which
differ for $0 \leq z < z_{\rm max}$ and $z > z_{\rm max}$,
are not especially illuminating and so we omit them here.

Self-gravity is ignored. Its neglect is justified for densities
less than the Roche density, $(\rho_g + \rho_d) \ll M/(2\pi r^3)$.
Equivalently, for standard nebular parameters, we are restricted to
$\mu \ll 30$. Sekiya (1998;
see also Youdin \& Shu 2002) relaxes this condition and
accounts for vertical self-gravity
in computing the Richardson number.

\begin{figure} 
\epsscale{1.2}
\plotone{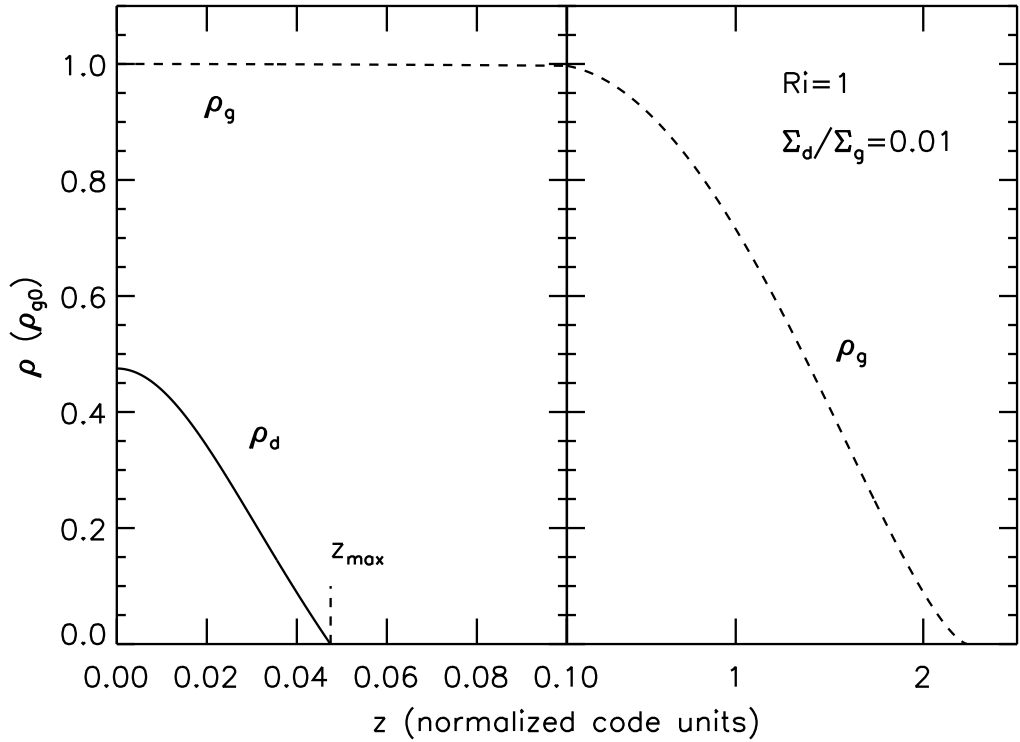}
\caption{Dust and gas vertical density profiles,
computed for constant $Ri=1$ and
vertical hydrostatic equilibrium. Disk self-gravity is neglected
(cf.~Sekiya 1998). 
Note the change in horizontal scale between the two panels.
The dust density $\rho_d$ is zero at $z=z_{\rm max}$.
Typically our simulations span $z = \pm 2 z_{\rm max}$;
the gas density $\rho_g$ is practically constant on these scales.}
\label{dens_1}
\end{figure}

\begin{figure} 
\epsscale{1.2}
\plotone{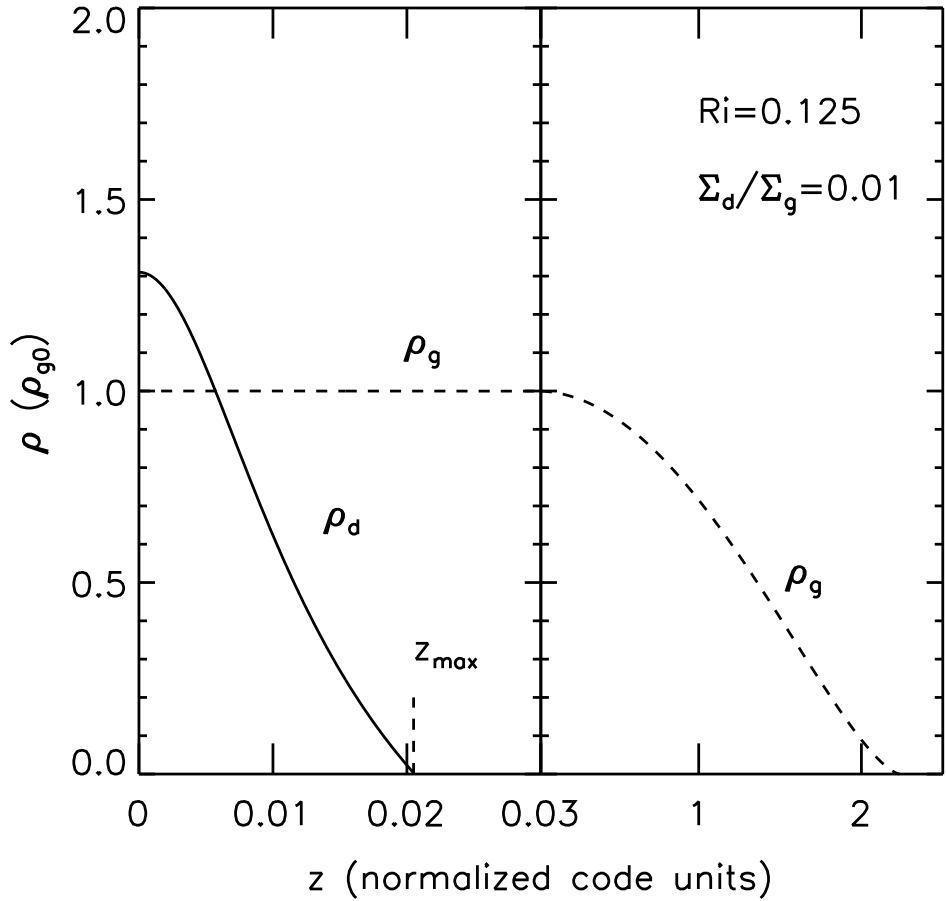}
\caption{Same as Figure \ref{dens_1}, but for $Ri = 0.125$.}
\label{dens_0.125}
\end{figure}

\begin{figure} 
\epsscale{1.2}
\plotone{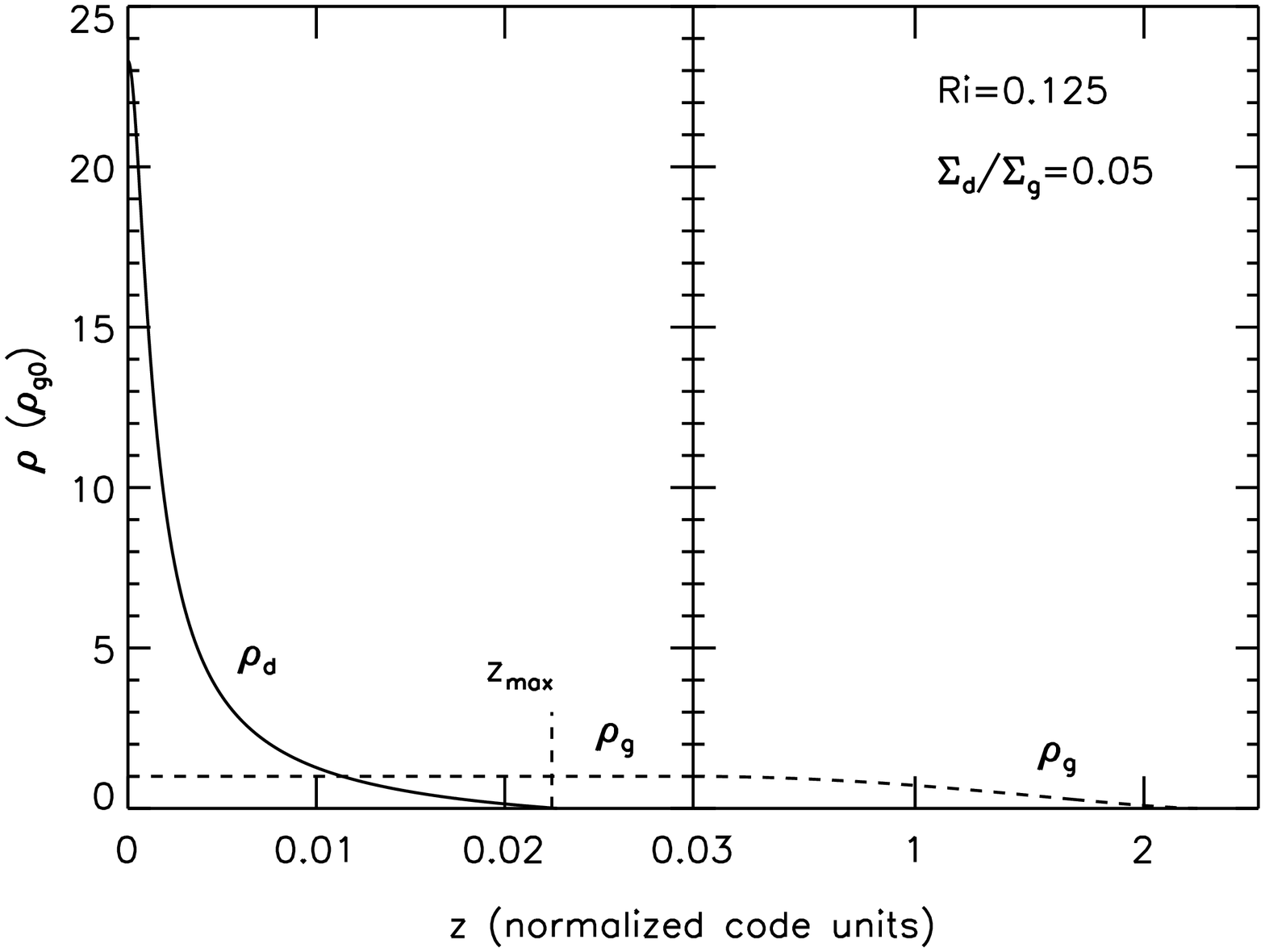}
\caption{Same as Figure \ref{dens_0.125}, but for a super-solar
metallicity $\Sigma_d/\Sigma_g=0.05$. The Richardson number, 0.125, is about
the lowest it can be without the dust layer
turning over for this metallicity (see
Run S8 of Table \ref{tab_shear}). The midplane dust density,
more than 20 times higher than the local gas density, is sufficiently high
that dust self-gravity can be dynamically important. Ways of achieving
$\Sigma_d/\Sigma_g = 0.05$ are mentioned in \S\ref{sum}.}
\label{dens_sigma05}
\end{figure}

Figure \ref{dens_1} shows profiles for $\rho_g(z)$ and $\rho_d(z)$,
for normalized (code) parameters of $\rho_{g0} = \rho_g(z=0) = 1$,
$K=1$, $\gamma=5/3$, $\Omega_0=1$, $Ri=1$, $\mu_0=0.475$, and
$v_{\rm max}/c_{s0}=0.05$, where
$c_{s0} = \sqrt{\gamma K \rho_{g0}^{\gamma-1}} = 1.3$
is the sound speed at the midplane. These parameter choices
imply a height-integrated dust-to-gas surface density ratio
(height-integrated metallicity) of $\Sigma_d / \Sigma_g =0.01$,
the nominal solar value.
We vary $Ri$ and $\mu_0$ from simulation to simulation.
Note that our choice of $v_{\rm max}/c_{s0} = 0.05$ is a factor
of 2 smaller than that of GO because we prefer to use
gas temperatures characterizing passive disks (see, e.g.,
Chiang et al.~2001) rather than the hotter temperatures of
the Hayashi (1981) nebula. Disk gas must be fairly passive, i.e., not
heated by turbulent dissipation, for dust to settle into the thin
layers of interest here (see Appendix B of Youdin \& Chiang 2004
for estimates of the degree of turbulence permitted).
A factor of 2 decrease in $v_{\rm max}/c_{s0}$ produces a factor
of 2 increase in the central concentration $\mu_0$, at fixed
$Ri$ and fixed $\Sigma_d / \Sigma_g$.
Figure \ref{dens_0.125} is analogous to Figure 1, except that
$Ri=0.125$ and $\mu_0 = 1.31$ (so that again $\Sigma_d/\Sigma_g=0.01$).
Figure \ref{dens_sigma05} is analogous to Figure 2, except
that $\mu_0 = 23.3$, so that $\Sigma_d/\Sigma_g = 0.05$
(the height-integrated metallicity is super-solar by a factor of 5).

To summarize this subsection, our equilibrium initial conditions
specify $v_y$ through (\ref{vy}), $\mu \equiv \rho_d/\rho_g$ through
(\ref{mu}), and $\rho_g$ through analytic solution of (\ref{hydro}).
Initially, $v_x = v_z = 0$, and $\rho_g$ and $\rho_d$ are constant
with $x$ and $y$. The primary input parameters are $Ri$ and
$\mu_0$; we hold $v_{\rm max}/c_{s0} = 0.05$ fixed for all runs.

\placefigure{fig1}

\placefigure{fig2}

\placefigure{fig3}

\subsection{Code}
\label{code}

We employ the hydrodynamics code ZEUS (Stone \& Norman 1992), a version
of which was kindly given to us by Dr.~Bryan Johnson. Because we
are interested in flows that shear both in radius ($\partial v_y / \partial
x$)
and height ($\partial v_y / \partial z$), the simulations
are necessarily three-dimensional. We adopt the usual shearing box (e.g.,
Hawley, Gammie, \& Balbus 1995), with
shearing periodic boundary conditions in $x$,
periodic boundary conditions in $y$, and closed boundary
conditions in $z$.

Three modifications are made to the code. The first is the addition
of another fluid, dust, which is transported according to
its own continuity equation (\ref{nine}), but otherwise has the same velocity
as that of gas. The second modification is to the pressure gradient source
term: $\nabla P / \rho_g \rightarrow \nabla P / (\rho_g + \rho_d)$,
in accordance with (\ref{abc}). The final change is to introduce
an extra source term for the ``large-scale'' radial pressure gradient,
distinct
from the just mentioned ``local'' pressure gradient. That is, equation
(\ref{a}) for the $x$-momentum is revised to read

\begin{equation}
\frac{\partial v_x}{\partial t} + v_i \frac{\partial v_x}{\partial x_i} =
\frac{-1}{\rho_g+\rho_d}\frac{\partial P}{\partial x} + 2\Omega_0 v_y + 2q
\Omega_0^2 x - \frac{1}{\rho_g+\rho_d} \left( \frac{\partial P}{\partial r}
\right)_{t=0} \,. \label{amod}
\end{equation}

\ni The extra source term
$(\partial P / \partial r)_{t=0}$ is given by
(\ref{dpdr0}). It accounts for how the pressure changes over radial
lengthscales of $r$, and is held fixed for each simulation.
Its inclusion is necessary to have dust-rich gas rotate faster
than dust-poor gas initially, i.e., to drive the vertical shear.
But because our simulations use a shearing box---in other words,
because they are local---we cannot follow the evolution
of this large-scale gradient. A self-consistent treatment
would require us to model the disk globally. Our use of an
external, constant, radial pressure gradient is the same device
employed by GO. The error introduced is
negligible insofar as changes in pressure over lengthscales of $r$ occur
over timescales much longer than the dynamical timescales which concern us
here.

In short, the modified code solves equations (\ref{eos})--(\ref{nine}),
(\ref{b}), (\ref{c}), and (\ref{amod}).
Initial conditions for the code are given above in \S\ref{init}.
The equilibrium state is perturbed by adding a random velocity
to each grid cell. The perturbation velocity points in a
random direction from cell to cell. For our main set of runs,
the magnitude of the perturbation in each cell
is drawn at random from a uniform distribution that extends from zero to
$10^{-3}c_{s0}$ (for comparison,
the maximum vertical shearing velocity is about $v_{\rm max} = 0.05 c_{s0}$).
Several auxiliary runs use much smaller or more spatially limited
initial perturbations; these are described
in \S\ref{sivp} and produce the same outcomes as our main simulations.

Our standard box dimensions are $L_x = L_y = 8 z_{\rm max}$
and $L_z = 4 z_{\rm max}$, and the corresponding number of grid
cells is $(N_x, N_y, N_z) = (64,64,32)$. The duration
of each simulation is at least $t_f = 20 \Omega_0^{-1}$.
Run parameters are contained in Tables 1 and 2, and are
justified in \S\ref{res}.

Including the von
Neumann \& Richtmyer (1950) artificial viscosity makes little difference
to our results---not surprisingly since our flows are highly
subsonic ($v_{\rm max}/c_{s0} = 0.05$). Therefore the simulations reported
in this paper have no artificially imposed viscosity apart from
the unavoidable numerical kind.

\section{RESULTS}
\label{res}

\subsection{Coriolis Only: $q=0$}
\label{cori}

To test our code, we repeat the experiments of GO. The Coriolis
acceleration is retained but the radial shear is suppressed
by setting $q=10^{-6}$ in the code. Effectively, this converts
shearing boundary conditions in $x$ to periodic boundary conditions.
It also
eliminates the contribution, $2q\Omega_0^2x$,
to $d{v}_x/dt$ from the centrifugal
force and tidal gravity.
Since the resultant flows have no structure in $x$, we reduce
$N_x$ to 2 and $L_x$ to $0.25 z_{\rm max}$.

\begin{figure} 
\epsscale{0.9}
\plotone{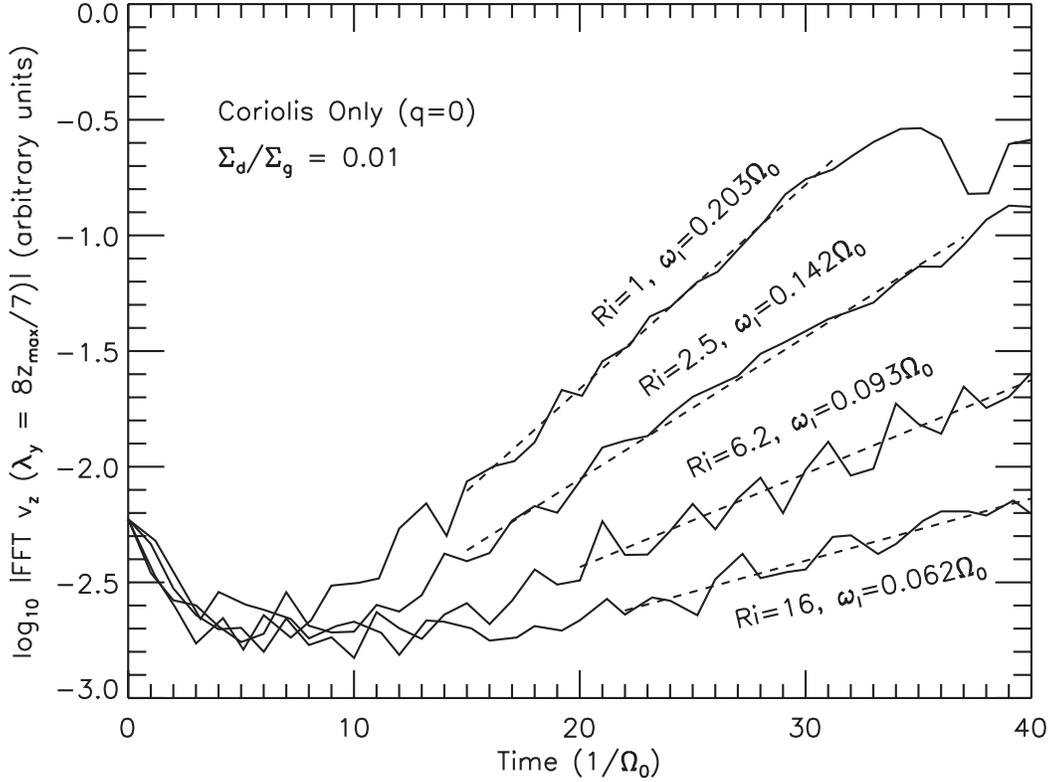} 
\caption{Fourier amplitudes versus time for the $\lambda_y/z_{\rm max}=8/7$
mode,
in simulations of varying $Ri$. At fixed $x = 0.0625 z_{\rm max}$, and given
$z$ and $t$,
we compute the fast Fourier transform (FFT) of
$v_z$ in the $y$-direction. What is plotted at a given $t$ is the maximum
value of the FFT (for the chosen $\lambda_y$) over all $z$. Growth rates
$\omega_I$ are the
slopes of lines fitted to the linear growth phase.
}
\label{f4}
\end{figure}

Table \ref{tab_cor} lists the run parameters. The main parameter
varied from run to run is $Ri$;
the central concentration $\mu_0$ is adjusted in tandem
to keep $\Sigma_d / \Sigma_g = 0.01$ fixed for all runs (this restriction
to solar metallicity
is relaxed for the fully shearing runs of \S\ref{cor+shear}).
Following GO, we diagnose the flow by Fourier analyzing $v_z$
in the $y$-direction at fixed $x$, $z$, and $t$. We then inspect
how the Fourier amplitudes change with time. We are interested in those
modes whose $y$-wavelengths $\lambda_y$ are comparable to the dust layer
thickness $z_{\rm max}$, as shorter wavelength modes cannot overturn
the layer and longer wavelength modes grow more slowly.
We are able to measure the exponential growth
rates $\omega_I$ of several modes whose $\lambda_y$'s range
from $\sim$8/15 to $\sim$8/3 of $z_{\rm max}$.\footnote{Periodic boundary
conditions in $y$ imply that the only modes present are those for which
the simulation box fits an integral number of wavelengths. 
The standard box length is $L_y = 8z_{\rm max}$.} Figure \ref{f4} displays
the Fourier amplitudes versus time for the $\lambda_y/z_{\rm max} = 8/7$ mode,
in simulations of varying $Ri$. While the growth rate $\omega_I$ of
this mode decreases
with increasing $Ri$, growth is still reliably detected
for $Ri$ as high as 16, a value $64 \times$ greater than the
canonical threshold of 1/4. Our measured growth rates are similar to
those reported by GO (see their Figure 10).

Runs C1a--C1f test the sensitivity of our results to simulation box size
and spatial resolution. Doubling the box height $L_z$ to $8z_{\rm max}$
at fixed resolution produces negligible change in measured growth rates
(C1a vs.~C1). Doubling the vertical resolution by doubling
$N_z$ at fixed $L_z$ increases mode growth rates by $\sim$10\% (C1b vs.~C1).
Changes in $\omega_I$ of order 10\% are also produced by varying
the azimuthal parameters $L_y$ and $N_y$ by factors of 2 (C1c--C1f). We
conclude
that our standard box size $(L_y$,$L_z) = (8,4)z_{\rm max}$
and resolution $(N_y$,$N_z)=(64,32)$ adequately balance the
need for accuracy with the need for computational speed.

\begin{figure}
\epsscale{0.9}
\plotone{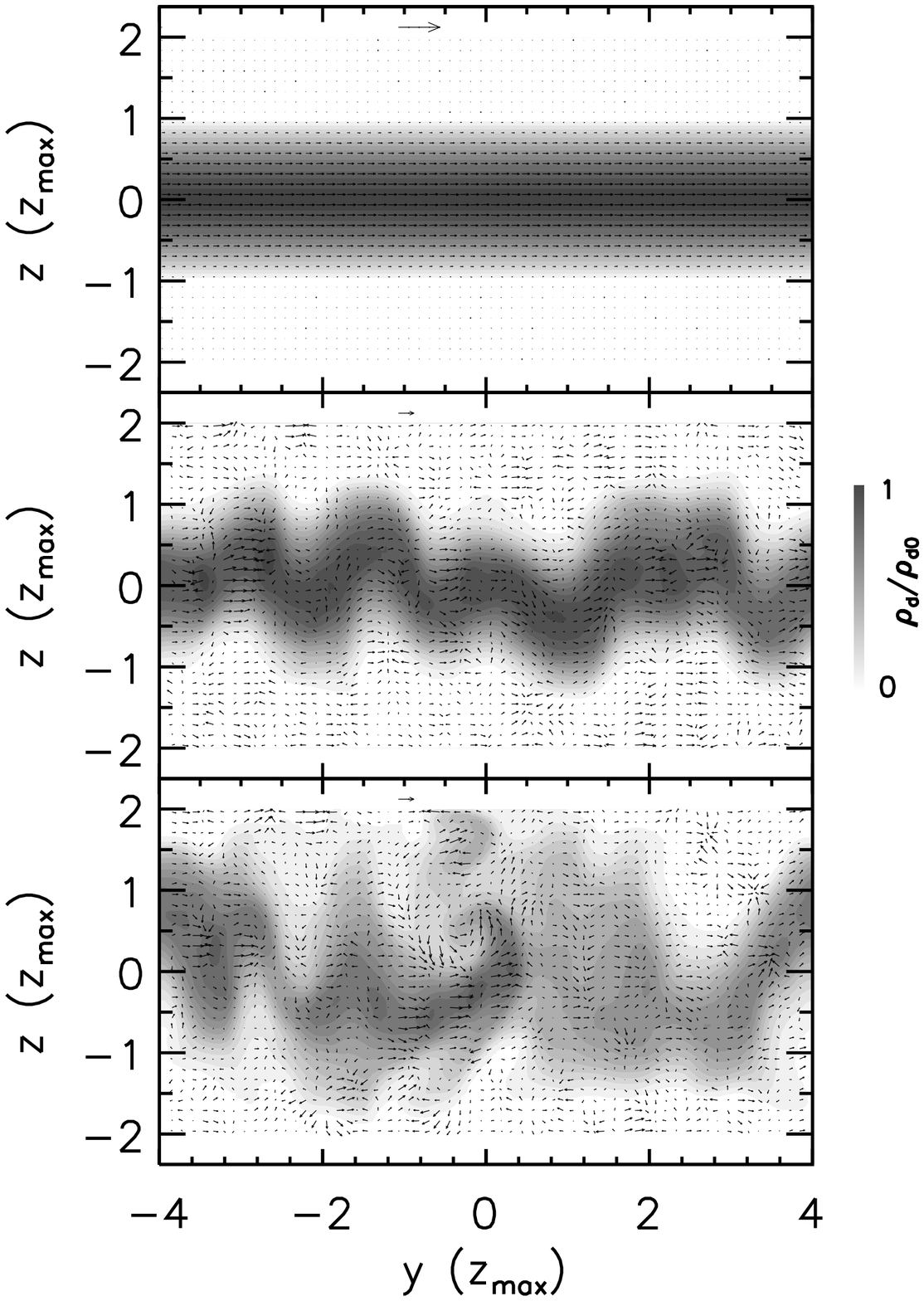} 
\vspace{-1in}
\caption{Snapshots of the $(y,z)$
flow for $Ri = 3.952$ and $\mu_0 = 0.270$,
including only Coriolis forces ($q=0$), at times $t=0$ (top), $t=95
\Omega_0^{-1}$ (middle),
and $t=170 \Omega_0^{-1}$ (bottom).
Greyscale denotes the density of dust only, normalized to the
density of dust at the midplane at $t=0$. Velocity vectors are
shown in the frame rotating with dust-free gas (see top panel). The lone arrow
at the top of each panel gives the length of $v_{\rm max}$ for that panel
only.
Data are based on a simulation for which $(L_x,L_y,L_z)=(0.25,8,4)z_{\rm max}$
and $(N_x,N_y,N_z) = (2,128,64)$.
}
\label{f5}
\end{figure}

Table \ref{tab_cor} and Figure \ref{f4} suggest that when only the Coriolis
force is included, and radial shear is omitted,
dust layers characterized by surprisingly large values for $Ri$
will eventually overturn and mix. This conclusion is supported by
Figure \ref{f5},
which displays snapshots of a high-resolution $(N_y,N_z)=(128,64)$
run for which $Ri = 4$.
Despite $Ri$ being 16 $\times$
greater than the traditional critical value of $1/4$,
the dust layer is clearly unstable. The fastest
growing mode has $\lambda_y/z_{\rm max} \approx 8/5$
and overturns the layer by $t \approx 100 \Omega_0^{-1}$.
All these results agree with those of GO.

We suspect that the instability discussed in this section, and simulated
first by GO, is a baroclinic instability afflicting $q=0$ disks.
Cabot (1984) finds that baroclinic, $q=0$ disks are linearly unstable
to non-axisymmetric perturbations;
that the ``baroclinic instability draws its energy from radial excursions by
fluid elements'' (radial excursions made possible by the Coriolis force);
and that maximum growth rates $\omega_I$ are of order
the maximum vertical shearing frequency
$\max |\partial v_y/\partial z|$ (see his Table 1 and
the discussion following his equation 22b).
For our dusty layers, $\max |\partial v_y/\partial z| \approx
 \Omega / \sqrt{Ri}$;
support for the inverse square-root dependence on $Ri$
can indeed be found in Figure \ref{f4}.\footnote{
Our shearing rates
are much larger than those considered by Cabot (1984),
who analyzed pure gas disks with no gradients in mean molecular weight.
When $\Sigma_d = 0$, the maximum vertical shear is
$\max |\partial v_y / \partial z| = \xi \Omega h / r$,
where $h \approx c_s/\Omega < r$ is the hydrostatic disk thickness
and $\xi$ is a number typically of order 0.1. The smallness of $\xi$ arises
because two effects compete
and nearly cancel: both the radial gravity $-\partial \Phi / \partial r$ and
the radial pressure acceleration $-(1/\rho) \partial P / \partial r$ decrease
in magnitude with increasing height.}

Knobloch \& Spruit (1986) extend the work of Cabot (1984) by restoring
the effects of Keplerian shear to baroclinic disks. They find that
radial shear tends to stabilize the flow to non-axisymmetric disturbances.
For their particular choice
for the form of $v_y(z) \propto -z^4$, the linear baroclinic instability
is defeated
when baroclinity is small ($|\partial v_y / \partial z| \ll \Omega$)
and when the disk is strongly stable to thermal convection in the vertical
direction. These stability requirements are usually satisfied in disks without
gradients in mean molecular weight (such disks are their main concern).
Knobloch \& Spruit (1986) speculate
that instability occurs when baroclinity is large.
In the context of their disks, large baroclinity demands
variations in specific entropy on short radial
length scales $\lesssim h$, the vertical height of disk
gas. Rapid radial variations are
also required for instability by Arlt \& Urpin (2004), who
study axisymmetric perturbations.

But large baroclinity
can also be achieved by having a strong molecular weight
gradient in the vertical direction, as exists in a highly settled
dust layer. We turn now to simulating such highly baroclinic,
radially shearing flows.

\begin{deluxetable}{ccccccccccccc}
\tablecaption{Coriolis Only ($q=0$) Simulations}
\tablewidth{0pt}
\tablehead{
\multicolumn{13}{c}{}\\
Name & $N_x$ & $L_x$ & $N_y$ & $L_y$ & $N_z$ & $L_z$ & $Ri$ & $\mu_0$ &
$\Sigma_d/\Sigma_g$ & $t_f$ &  $\lambda_y$$^a$ & $\omega_I$ \\
     &       & ($z_{\rm max}$) & & ($z_{\rm max}$) & & ($z_{\rm max}$) & & & &
($\Omega_0^{-1}$) & ($z_{\rm max}$) & ($\Omega_0$) \\
}
\startdata
C1a$^b$  & 2 & 0.25 & 64 & 8 & 64 & 8 & 0.25  & 0.903 & 0.01 & 20 & 8/3 & 0.334
\\
C1b & '' & '' & 64 & 8 & 64 & 4 & ''  & '' & '' & '' & 8/3 & 0.371 \\
C1c & '' & '' & 64 & 16/3 & 32 & 4 & '' & '' & '' & '' & 8/3 & 0.352 \\
C1d & '' & '' & 32 & 8/3 & '' & '' & ''  & '' & '' & '' & 8/3 & 0.368 \\
C1e & '' & '' & 32 & 3   & '' & '' & ''  & '' & '' & '' & 3   & 0.339 \\
C1f & '' & '' & 64 & 7   & '' & '' & ''  & '' & '' & '' & 7/3 & 0.336 \\
C1 & '' & '' & 64 & 8 & 32 & 4 & ''  & '' & ''  & 40 & 8/3 & 0.332 \\
    &   &      &    &   &    &   &       &       &      &    & 8/5 & 0.303 \\
    &   &      &    &   &    &   &       &       &      &    & 8/7 & 0.342 \\
\hline
C2 & '' & '' & '' & '' & '' & '' & 1     & 0.475 & '' & '' & 8/3 & NA$^c$
\\
    &   &      &    &   &    &   &       &       &      &  & 8/5 & 0.165 \\
    &   &      &    &   &    &   &       &       &      &  & 8/7 & 0.203 \\
C3 & '' & '' & '' & '' & '' & '' & 1.581 & 0.391 & '' & '' & 8/3 & NA    \\
    &   &      &    &   &    &   &       &       &      &  & 8/5 & 0.148 \\
    &   &      &    &   &    &   &       &       &      &  & 8/7 & 0.171 \\
C4 & '' & '' & '' & '' & '' & '' & 2.5   & 0.324 & '' & '' & 8/3 & NA    \\
    &   &      &    &   &    &   &       &       &      &  & 8/5 & 0.088 \\
    &   &      &    &   &    &   &       &       &      &  & 8/7 & 0.142 \\
C5 & '' & '' & '' & '' & '' & '' & 3.952 & 0.270 & '' & '' & 8/3 & NA    \\
    &   &      &    &   &    &   &       &       &      &  & 8/5 & 0.084 \\
    &   &      &    &   &    &   &       &       &      &  & 8/7 & 0.125 \\
C6 & '' & '' & '' & '' & '' & '' & 6.248 & 0.226 & '' & '' & 8/3 & NA    \\
    &   &      &    &   &    &   &       &       &      &  & 8/5 & NA    \\
    &   &      &    &   &    &   &       &       &      &  & 8/7 & 0.093 \\
C7 & '' & '' & '' & '' & '' & '' & 9.878 & 0.190 & '' & '' & 8/3 & NA    \\
    &   &      &    &   &    &   &       &       &      &  & 8/5 & NA    \\
    &   &      &    &   &    &   &       &       &      &  & 8/7 & 0.078 \\
C8 & '' & '' & '' & '' & '' & '' & 15.62 & 0.161 & '' & '' & 8/3 & NA    \\
    &   &      &    &   &    &   &       &       &      &  & 8/5 & NA    \\
    &   &      &    &   &    &   &       &       &      &  & 8/7 & 0.062 \\
\enddata
\tablenotetext{a}{Wavelength of a sampled mode. Other modes exist and were
measured that
are not listed here. In runs C1a--C1f, the listed mode is the strongest mode.}
\tablenotetext{b}{Runs C1a--C1f experiment with the size and spatial resolution
of the simulation box.}
\tablenotetext{c}{NA denotes a mode whose Fourier amplitude does not grow
smoothly and exponentially over
the duration of the simulation. In most cases the amplitude either remains
roughly constant
or decreases.
}
\label{tab_cor}
\end{deluxetable}

\placefigure{fig4}

\placefigure{fig5}

\subsection{Coriolis + Radial Shear: $q=3/2$}
\label{cor+shear}

Having found simulation parameters $(L_y,L_z,N_y,N_z)$ that
produce realistic results for $q=0$, we use those same
parameters for our radially shearing $q=3/2$ simulations.
Since the radial wavenumber of a disturbance grows at rate
$\dot{k}_x = q \Omega_0 k_y$, our resolution in $x$ should be
at least comparable to that in $y$. Our standard run parameters
are $(L_x,L_y,L_z) = (8,8,4)z_{\rm max}$ and $(N_x,N_y,N_z)=(64,64,32)$.
Table \ref{tab_shear} lists the various experiments, all of which start with
$Ri =$ constant flow as described in \S\ref{init}. Whether
the dust layer turns over within the run duration is indicated
in the table.

\begin{deluxetable}{ccccccccccccc}
\tablecaption{Fully Shearing ($q=3/2$) Simulations}
\tabletypesize{\small}
\tablewidth{0pt}
\tablehead{
\multicolumn{13}{c}{}\\
Name & $N_x$ & $L_x$ & $N_y$ & $L_y$ & $N_z$ & $L_z$ & $Ri$ & $\mu_0$ &
$\Sigma_d/\Sigma_g$ & Noise$^a$ & $t_f$ &  Turns \\
     &       & ($z_{\rm max}$) & & ($z_{\rm max}$) & & ($z_{\rm max}$) & & & &
$(c_{s0})$ & ($\Omega_0^{-1}$) & Over? \\
}
\startdata
S1  & 64  & 8  & 64  & 8  & 32 & 4  & 0.25    & 0.903 & 0.01 & $10^{-3}$ & 40 &
N$^b$ \\
S2  & ''  & '' & ''  & '' & '' & '' & 0.125   & 1.31  & ''   & $10^{-3}$ & 100
& N \\
S2a & 128 & '' & 128 & '' & 64 & '' & ''      & ''    & ''   & $10^{-3}$ & 20 &
N \\
S3  & 64  & 8  & 64  & 8  & 32 & 4  & 0.0625  & 1.989 & ''   & $10^{-3}$ & 50 &
Y \\
S3a & ''  & '' & ''  & '' & '' & '' & ''      & ''    & ''   & $10^{-6}$ & 70 &
Y \\
S4  & ''  & '' & ''  & '' & '' & '' & 0.03125 & 3.252 & ''   & $10^{-3}$ & 20 &
Y \\
S4a & ''  & '' & ''  & '' & '' & '' & ''      & ''    & ''   & $10^{-6}$ & 40 &
Y \\
\hline
S5  & ''  & '' & ''  & '' & ''     & '' & 0.25    & 3.57 & 0.03 & $10^{-3}$ &
20 & N \\
S6  & ''  & '' & ''  & '' & 48$^c$ & '' & 0.125   & 6.65 & ''   & $10^{-3}$ &
20 & N \\
S7  & ''  & '' & ''  & '' & 64$^c$ & '' & 0.0625  & 14.7 & ''   & $10^{-3}$ &
20 & Y \\
\hline
S8$^d$  & ''  & '' & '' & '' & 64$^c$ & '' & 0.125 & 23.3 & 0.05 & $10^{-3}$ &
20 & N \\
\hline
S9  & ''  & '' & ''  & '' & 32 & '' & 0.125   & 0.174 & 0.001 & $10^{-3}$ & 70
& N \\
S10 & ''  & '' & ''  & '' & '' & '' & 0.0625  & 0.226 & ''   & $10^{-3}$ & 70 &
N \\
S11 & ''  & '' & ''  & '' & '' & '' & 0.03125 & 0.296 & ''   & $10^{-3}$ & 70 &
N \\
S12 & ''  & '' & ''  & '' & '' & '' & 0.01563 & 0.393 & ''   & $10^{-3}$ & 70 &
Y \\
\enddata
\tablenotetext{a}{Maximum initial perturbation velocity applied to each grid
cell. From cell to cell, the
perturbation velocity points in a random direction, and its magnitude
is drawn randomly
from a uniform distribution from 0 to the maximum indicated in the table.
For comparison, the maximum vertical shearing velocity is about $v_{\rm
max}=0.05 c_{s0}$.}
\tablenotetext{b}{For runs in which the layer does not overturn, we
perform the same Fourier analysis that we do for Coriolis-only
runs, and detect no growing modes at all.}
\tablenotetext{c}{Increasing $\mu_0$ (equivalently, $\Sigma_d / \Sigma_g$)
at fixed $Ri$ steepens the dust density profile $\rho_d(z)$, whose
resolution then requires greater $N_z$.}
\tablenotetext{d}{We do not simulate
$Ri = 0.0625$ and $\Sigma_d/\Sigma_g = 0.05$,
since for such parameters $\mu_0 = 79.5$; dust self-gravity, which we do
not account for, would be significant in this case.
}
\label{tab_shear}
\end{deluxetable}

\begin{figure}
\epsscale{0.9}
\plotone{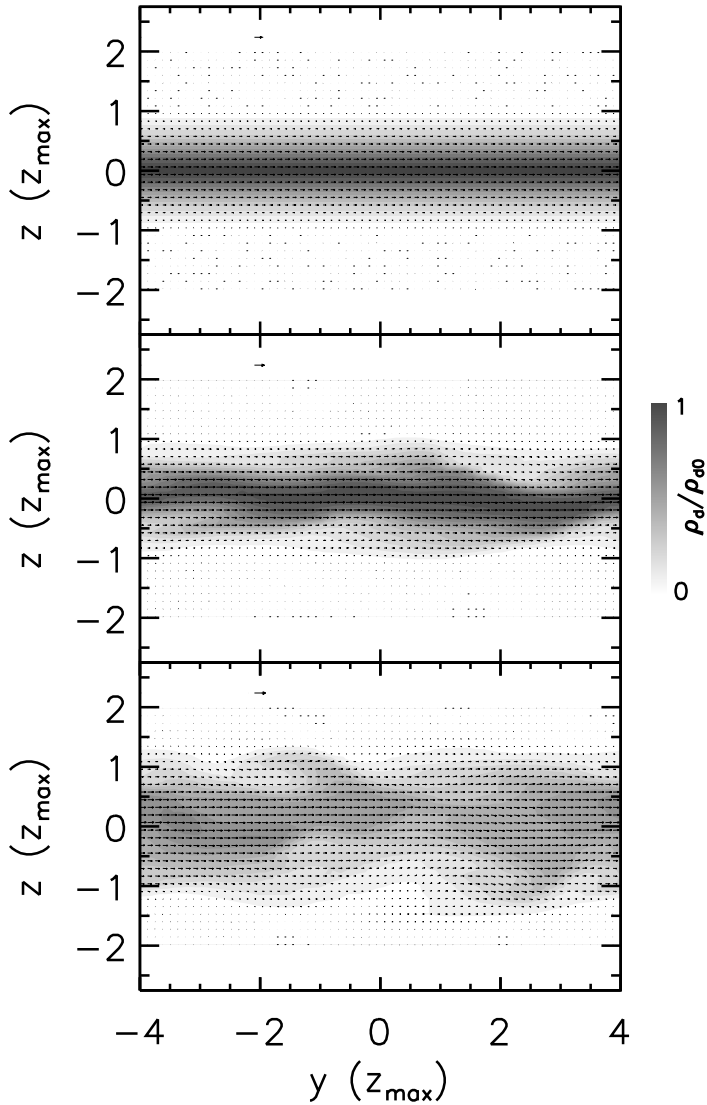} 
\vspace{-1in}
\caption{Snapshots of the $(y,z)$
flow, at fixed $x=0$, for $Ri = 0.0625$ and $\mu_0 = 1.989$,
in a fully shearing box ($q=3/2$),
at times $t=0$ (top), $t=24 \Omega_0^{-1}$ (middle),
and $t=37 \Omega_0^{-1}$ (bottom).
Greyscale denotes the density of dust only, normalized to the
density of dust at the midplane at $t=0$. Velocity vectors are
shown in the frame rotating with dust-free gas (see top panel). The lone arrow
at the top of each panel gives the length of $v_{\rm max}$ for that panel
only.
Data are from simulation S3 for which $(L_x,L_y,L_z)=(8,8,4)z_{\rm max}$,
$(N_x,N_y,N_z) = (64,64,32)$, and $\Sigma_d/\Sigma_g=0.01$.
}
\label{f6}
\end{figure}

\begin{figure}
\epsscale{0.9}
\plotone{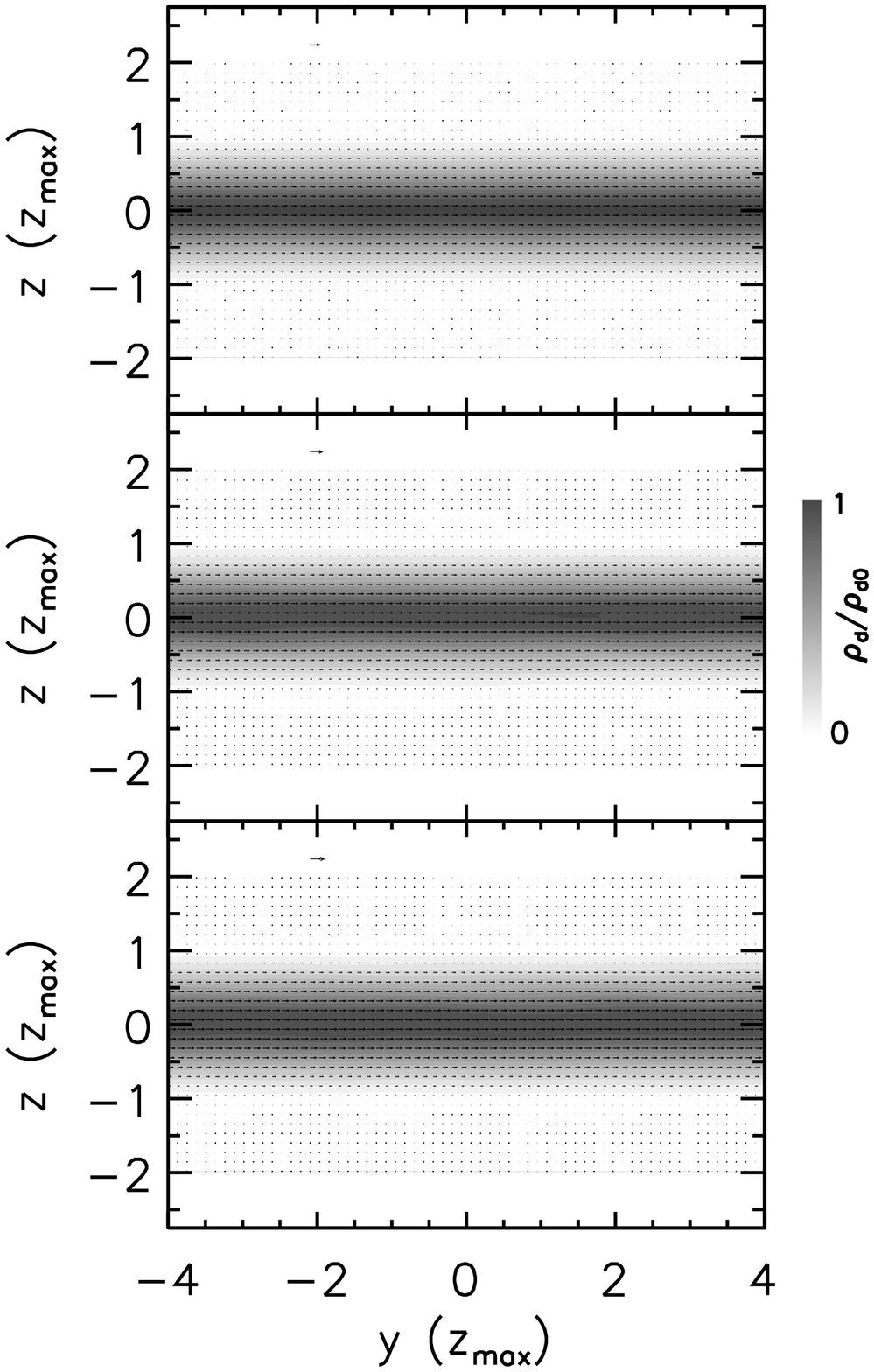} 
\vspace{-1in}
\caption{Snapshots of the $(y,z)$ flow, at fixed $x=0$,
for $Ri = 0.125$ and $\mu_0 = 1.31$,
in a fully shearing box ($q=3/2$),
at times $t=0$ (top), $t=25 \Omega_0^{-1}$ (middle),
and $t=50 \Omega_0^{-1}$ (bottom).
Greyscale denotes the density of dust only, normalized to the
density of dust at the midplane at $t=0$. Velocity vectors are
shown in the frame rotating with dust-free gas (see top panel). The lone arrow
at the top of each panel gives the length of $v_{\rm max}$ for that
panel only.
Data are from simulation S2 for which $(L_x,L_y,L_z)=(8,8,4)z_{\rm max}$,
$(N_x,N_y,N_z) = (64,64,32)$, and $\Sigma_d/\Sigma_g=0.01$.
}
\label{f7}
\end{figure}

Results for $q=3/2$ differ dramatically from those for $q=0$.
Whereas instability characterizes all values of $Ri$ for $q=0$,
we find that $Ri$ must fall below a critical value, $Ri_{\rm crit} \approx
0.1$,
for the layer to overturn when $q=3/2$ and when $\Sigma_d/\Sigma_g=0.01$
(see \S\ref{varsig} for experiments that vary $\Sigma_d/\Sigma_g$).
Figures \ref{f6} and \ref{f7}
demonstrate this point: the flow for $Ri=0.0625$ eventually mixes whereas that
for $Ri=0.125$ keeps the dust layer intact
for as long as $t_f = 100 \Omega_0^{-1}$.
The same Fourier analysis of \S\ref{cori}, applied to the latter run,
reveals no growth of any mode.
Repeating the $Ri=0.125$ run at
higher resolution---$(N_x,N_y,N_z)=(128,128,64)$---confirms these
results (run S2a).

\placefigure{fig6}

\placefigure{fig7}

\placefigure{fig8}
\begin{figure}
\epsscale{0.9}
\plotone{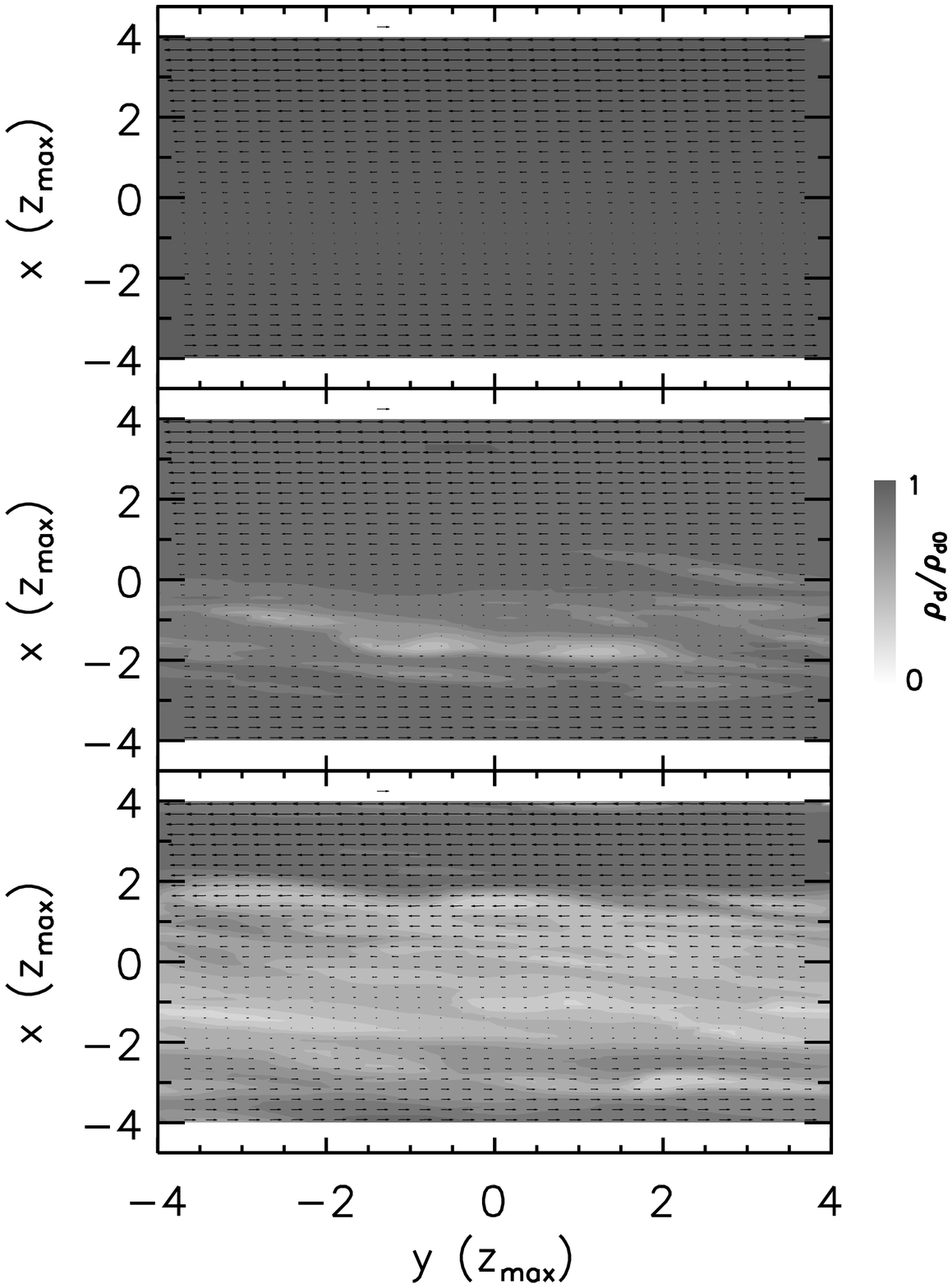} 
\vspace{-1in}
\caption{Snapshots of the midplane $(x,y)$ flow, at fixed $z=0$,
for $Ri = 0.0625$ and $\mu_0 = 1.989$,
in a fully shearing box ($q=3/2$),
at times $t=0$ (top), $t=24 \Omega_0^{-1}$ (middle),
and $t=37 \Omega_0^{-1}$ (bottom).
Greyscale denotes the density of dust only, normalized to the
density of dust at the midplane at $t=0$. Velocity vectors are
shown in the frame rotating at $\Omega_0$. Note how $v_y=0$ at $x < 0$,
not at the usual $x=0$,
a consequence of the imposed background radial pressure gradient
$(\partial P / \partial r)_{t=0}$ that enforces sub-Keplerian flow.
The lone arrow
at the top of each panel gives the length of $v_{\rm max}$
for that panel only.
Data are from simulation S3 for which $(L_x,L_y,L_z)=(8,8,4)z_{\rm max}$,
$(N_x,N_y,N_z) = (64,64,32)$, and $\Sigma_d/\Sigma_g=0.01$.
}
\label{f8}
\end{figure}

\subsubsection{Sensitivity to Initial Velocity Perturbations}\label{sivp}
Naturally, reducing the initial velocity perturbations
increases the time required for the layer to overturn.
The dependence is very slight; lowering the maximum
perturbation velocity from $10^{-3}c_{s0}$ to $10^{-6}c_{s0}$
increases the time to overturn by $10$--$20\Omega_0^{-1}$
(runs S3a and S4a).
(Nonetheless, eliminating the initial perturbations altogether
produces no evolution in the shearing box whatsoever.)

The instability
grows fastest at ``co-rotation,'' where $v_y = 0$, and spreads
radially inward and outward from that location. This behavior is evident in
Figure \ref{f8}, and we observe it in all our unstable runs.
Upon removing all initial velocity
perturbations in a narrow annulus surrounding co-rotation,
and leaving the perturbations in place everywhere else,
we find that the flow outside co-rotation still overturns, though more slowly.
Eventually the instability
spreads to co-rotation and mixes dust uniformly throughout
the entire box. This experiment assures us that the instability is not
an artifact of $v_y=0$. At present we attribute 
all this behavior to the fact that our fixed-amplitude
initial velocity perturbations
are fractionally larger near co-rotation; i.e.,
the initial perturbation $\delta v_y \sim 10^{-3} c_{s0}$
is a greater fraction of $v_y$ near co-rotation than elsewhere.
It would therefore be natural to expect the instability
to manifest itself most quickly at co-rotation.

\subsubsection{Varying the Height-Integrated Metallicity
$\Sigma_d/\Sigma_g$}\label{varsig}

For solar or super-solar values of $\Sigma_d/\Sigma_g = 0.01$--0.05,
the value of $Ri_{\rm crit}$ changes little from 0.1,
as runs S5--S8 of Table \ref{tab_shear} attest.
However, $Ri_{\rm crit}$ changes substantially
for sub-solar metallicities---it decreases to
$\sim$0.02 for $\Sigma_d/\Sigma_g = 0.001$ (runs S9--S12).
This result indicates that
the Richardson number alone is an inadequate predictor
of instability under general circumstances.
Still, because sub-solar metallicities seem less relevant
for planet formation, we have not pursued the question
of what should replace the Richardson criterion,
and we content ourselves with citing $Ri_{\rm crit} \approx 0.1$
with the understanding that this result applies only
for solar and moderately super-solar metallicities.

\begin{figure} 
\epsscale{1}
\plotone{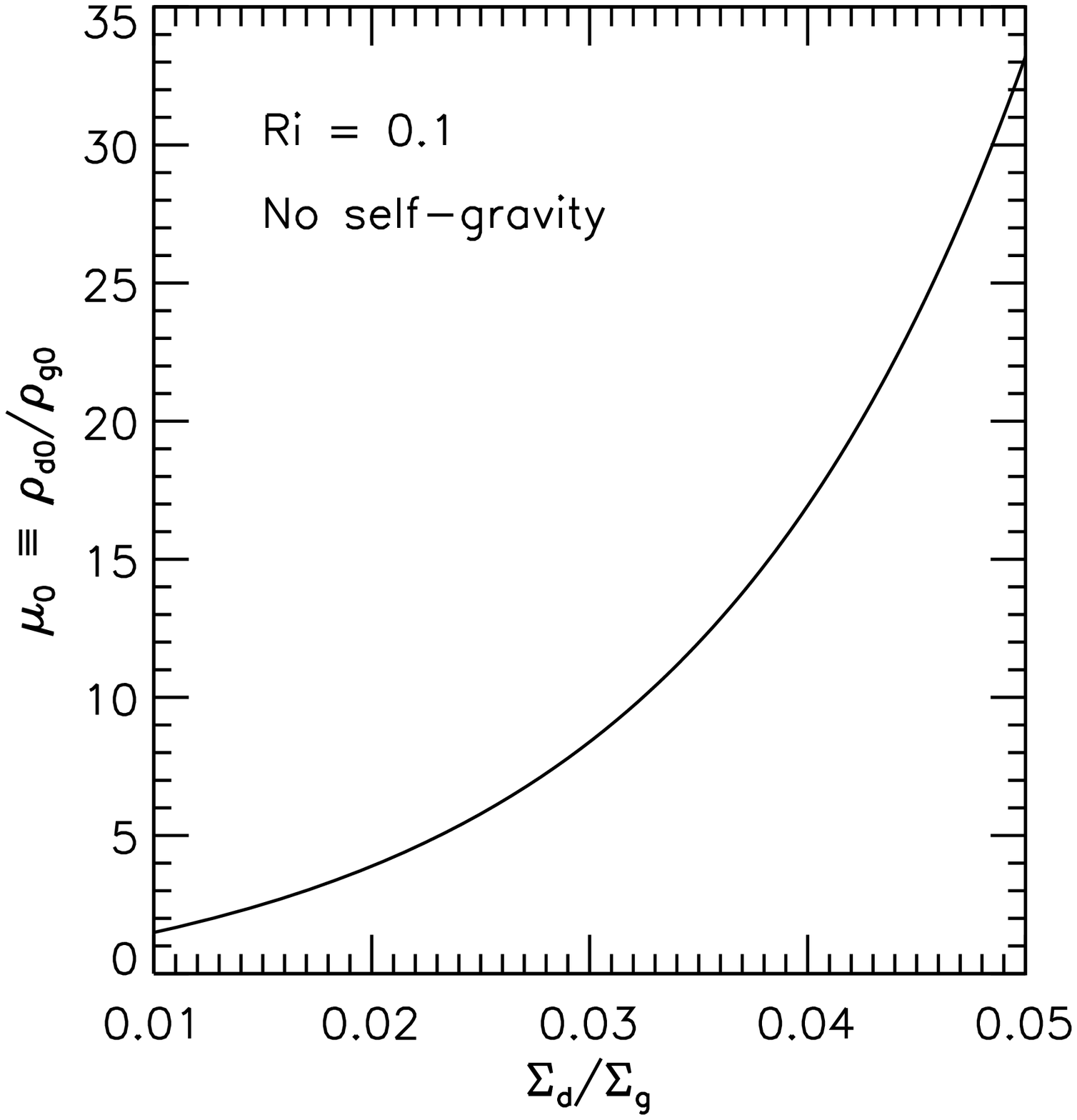}
\caption{One-to-one correspondence between the central concentration
$\mu_0$ and height-integrated metallicity $\Sigma_d/\Sigma_g$ at
fixed $Ri = 0.1 \approx Ri_{\rm crit}$. Once the height-integrated
metallicity is several times solar, the midplane dust-to-gas
density $(1+\mu_0)\rho_{g0}$ can approach Roche densities (see \S\ref{sum}).
}
\label{f9}
\end{figure}

Values of $\mu_0$ and $\Sigma_d/\Sigma_g$
at fixed $Ri = 0.1 \approx Ri_{\rm crit}$ are displayed in Figure \ref{f9}.
In a minimum-mass nebula for which
$\Sigma_d / \Sigma_g = 0.01$, $Ri = 0.1$ corresponds to a Toomre
$Q \approx M/[2\pi r^3 \rho_{g0} (1+\mu_0)] \approx 25$,
independent of disk radius if $\rho_{g0} \propto r^{-3}$, as
is approximately the case in standard nebular models.
Under these conditions, our neglect of the dust layer's self-gravity is
well justified.
Only when $\Sigma_d/\Sigma_g \gtrsim 0.05$ ($\mu_0 \gtrsim 30$; see
run S8)
does dust self-gravity become important (Sekiya 1998).

\subsubsection{Baroclinic vs.~Kelvin-Helmholtz}
We cannot say whether the observed turnover of the dust layer in a
fully shearing disk is better ascribed
to the Kelvin-Helmholtz instability (as studied by, e.g., Chandrasekhar 1961)
or to the baroclinic instability (as studied by Cabot 1984, Knobloch \&
Spruit 1986, and Arlt \& Urpin 2004).
Both instabilities rely on the vertical shear.
For $Ri\approx 0.1$, the maximum vertical shearing frequency
$\sim$$\Omega/\sqrt{Ri}$ exceeds, by factors of a few, both the maximum
Brunt-V\"ais\"al\"a frequency $\sim$$\Omega$ (thereby
satisfying the traditional criterion for the KHI) and the rotation frequency
$\Omega$ (thereby satisfying the criterion for ``large baroclinity,'' as
defined by Knobloch \& Spruit 1986).
An analytic criterion for the baroclinic instability, relevant
for non-axisymmetric perturbations, that is valid near the midplane
of a radially shearing disk and for arbitrary values of the vertical shear,
is not known (Knobloch \& Spruit 1986).

\placefigure{fig9}

\section{SUMMARY}
\label{sum}
We have performed numerical simulations of flows near the dust-rich midplanes
of protoplanetary disks. Dust particles are assumed small enough that they
are perfectly entrained in gas, and disk self-gravity is neglected.
The midplane flow is doubly shearing; the rotational velocity varies
with height according to the vertical gradient in dust density,
and it also varies with radius according to the usual Keplerian shear.
The simulations are necessarily three-dimensional and performed in a
shearing box.

Despite the complications introduced by rotation and tidal gravity,
the Richardson criterion for the onset of the Kelvin-Helmholtz
instability still proves useful for solar to moderately super-solar
height-integrated metallicities. Dust layers characterized by constant
Richardson number and $\Sigma_d/\Sigma_g \approx 0.01$--0.05
overturn and mix if $Ri \lesssim 0.1$, but remain
intact at larger $Ri$. This result contrasts with the situation when only
the Coriolis force is accounted for and the radial shear is suppressed:
in that case, unstable modes persist for $Ri$ as large
as 16. But because these ``Coriolis-only'' (and likely baroclinic;
Cabot 1984) modes at large $Ri$
grow at rates that are substantially slower than the Kepler strain
rate of $3\Omega/2$, they are stabilized and rendered impotent
by the Kepler shear.

In hindsight, our guess (\S\ref{intro}) that the critical $Ri$ dividing
stability from instability might only change by a factor of order unity
proved correct, though only for solar to moderately super-solar
metallicities: we find that $Ri_{\rm crit}$ for our doubly shearing
flows is about half that of the traditional value of 1/4,
if $\Sigma_d/\Sigma_g \approx 0.01$--0.05.
This is sensible insofar as $Ri_{\rm crit} \approx 0.1$ gives a vertical
shearing frequency $\partial v_y / \partial z$ that is a few times
faster than any of the other frequencies of the problem:
the Brunt-V\"{a}is\"{a}l\"{a} frequency for vertical oscillations
(this reaches a maximum of $\sim$$\Omega$ at $z\sim z_d$),
the Kepler strain rate ($3\Omega/2$), and the Coriolis
turning frequency ($2\Omega$). Deeper insights, including
an understanding of why $Ri_{\rm crit} \ll 0.1$
for $\Sigma_d/\Sigma_g \ll 0.01$, might be gained
by study of the geophysical literature, where doubly shearing,
rotating flows are commonplace (see, e.g., Chapter 7 of Pedlosky 1979,
which discusses baroclinic instabilities).

We conclude that for a
standard, height-integrated, solar metallicity of $\Sigma_d/\Sigma_g = 0.01$,
the dust density at the midplane can be at most $\sim$$1.5 \times$
greater than the gas density. If we assume a gas density appropriate
to a minimum-mass disk,
then the maximum total density of gas and dust is still too low, by a factor
of $\sim$25, for the marginally stable layer to become
gravitationally unstable. The literature discusses
two remedies. The first is to find ways
of enhancing $\Sigma_d/\Sigma_g$ by factors of 3--10, thereby
increasing the dust-to-gas ratio at the midplane by factors of $\sim$30
at fixed $Ri = 0.1$ (Sekiya 1998;
Youdin \& Shu 2002; see also our Figures \ref{dens_sigma05} and \ref{f9},
and run S8 in Table 2).
Possible means of increasing the height-integrated metallicity
include (i) decreasing $\Sigma_g$ through photoevaporation of gas,
(ii) increasing $\Sigma_d$ by radial drift and pile-up of particles
(Youdin \& Shu 2002; Youdin \& Chiang 2004) or (iii) increasing
$\Sigma_d$ by radiation blow-back
of grains into the rim of a transitional disk (Chiang \& Murray-Clay 2007).
Gas giants that form in such a metal-rich environment might be expected
to have metallicities enhanced above the solar value by similar
factors of 3--10.
Indeed some hot Jupiters have remarkably metal-rich interiors
(Sato et al.~2005).
Alternatively, one can relax the assumption of perfect
coupling between dust and gas, and exploit drag
instabilities that result from finite
momentum stopping times and the backreaction of dust on gas
(Youdin \& Goodman 2005; Johansen et al.~2007).

\acknowledgements
We thank Anders Johansen, Bryan Johnson, Eve Ostriker, Jack Wisdom, and
Andrew Youdin for helpful discussions. We owe to Jeremy Goodman
an insightful referee's report that brought to light the
baroclinic instability.
This work was supported by NSF grant AST-0507805.

\end{document}